\documentclass[%
 reprint,
 superscriptaddress,
 amsmath,amssymb,
aps,
pre
]{revtex4-1}

\usepackage{graphicx}
\usepackage{bm}
\usepackage{array}
\usepackage{tabularx}
\usepackage{dcolumn}
\newcolumntype{d}[0]{D{.}{.}{1}} 

\let\oldeqref\eqref
\renewcommand{\eqref}[1]{Eq.~\oldeqref{#1}}

\renewcommand{\b}[1]{\boldsymbol{\mathbf{#1}}}

\newcommand{\td}[2]{\frac{\mathrm{d} #1}{\mathrm{d} #2}}

\newcommand{\ii}[0]{\mathrm{i}}

\begin{document}


\title{Modeling synchronization in forced turbulent oscillator flows}

\author{Benjam\'in Herrmann}
\email{\vspace{-.2in}benherrm@uw.edu}
\affiliation{Department of Mechanical Engineering, University of Washington, Seattle, WA 98195, USA}
\affiliation{Institute of Fluid Mechanics, Technische Universit\"at Braunschweig, 38108 Braunschweig, Germany}
\author{Philipp Oswald}
\affiliation{Institute of Fluid Mechanics, Technische Universit\"at Braunschweig, 38108 Braunschweig, Germany}
\author{Richard Semaan}
\affiliation{Institute of Fluid Mechanics, Technische Universit\"at Braunschweig, 38108 Braunschweig, Germany}
\author{Steven L. Brunton}
\affiliation{Department of Mechanical Engineering, University of Washington, Seattle, WA 98195, USA}


\begin{abstract}
Periodically forced, oscillatory fluid flows have been the focus of intense research for decades due to their richness as a nonlinear dynamical system and their relevance to applications in transportation, aeronautics, and energy conversion. 
%
Recently, it has been observed that turbulent bluff-body wakes exhibit a subharmonic resonant response when excited with specific spatial symmetries at twice the natural vortex shedding frequency, which is hypothesized to be caused by triadic interactions. 
The focus of this paper is to provide new physical insight into the dynamics of turbulent oscillator flows, based on improved mechanistic models informed by a comprehensive experimental study of the turbulent wake behind a D-shaped body under periodic forcing. 
%
We confirm for the first time the role of resonant triadic interactions in the forced flow by studying the dominant components in the power spectra across multiple excitation frequencies and amplitudes. 
We then develop an extended Stuart-Landau model for the forced global wake mode, incorporating parametric and non-harmonic forcing.  
This model captures the system dynamics and reveals the boundaries of multiple synchronization regions.  
Further, it is possible to identify model coefficients from sparse measurement data, making it applicable to a wide range of turbulent oscillator flows.  
We believe these generalized synchronization models will be valuable for prediction, control, and understanding of the underlying physics in this ubiquitous class of flows.
\end{abstract}

\maketitle

\section{Introduction}
Fluid flows that display unsteadiness characterized by a well-defined frequency and that are insensitive to low-level external noise are known as oscillator flows~\citep{Huerre1990, Chomaz2005}. These flows have been the focus of research efforts for over 75 years~\cite{Landau1944,Stuart1960}, in part because of their rich physics, and also because of their relevance to numerous applications where aerodynamic forces and mixing play a significant role, such as transportation, aeronautics, and energy conversion~\citep{Brunton2015}.
Models that capture the evolution of dominant fluid coherent structures are of the utmost importance for prediction, control, and understanding of the underlying physical processes that drive these flows~\citep{Holmes2012}. The wake past a bluff body is one example of an oscillator flow, where self-sustained periodic vortex shedding arises after an increase in the Reynolds number renders the flow incapable of maintaining a steady state. This scenario unfolds when a supercritical Hopf bifurcation takes place, where disturbances associated with a spatial structure, known as the global mode of the flow, become linearly unstable, leading to exponential growth of the mode amplitude $A$, followed by nonlinear saturation onto a stable limit cycle~\citep{Sipp2007, Bagheri2013}. The Stuart-Landau model~\cite{Landau1944,Stuart1960}
\begin{equation}
\td{A}{t} = \sigma A - l|A|^2 A,\label{SL}
\end{equation}
\noindent
has been widely used to explain this nonlinear oscillator behavior of the wake past bluff bodies~\citep{Mathis1984, Noack2003, Thompson2004, Sipp2007, Bagheri2013, Gallaire2016}. 
Sipp and Lebedev~\cite{Sipp2007} formally derived this model from the Navier-Stokes equations by means of a rigorous asymptotic expansion close to the Hopf bifurcation.

When periodically forced with certain frequencies, 
the wake past a bluff body has been observed to adjust its natural vortex shedding rate to some rational multiple of the forcing frequency, as first reported by Provansal et al.~\cite{Provansal1987} for the cylinder flow. This is synchronization -- the spontaneous emergence of rhythmic oscillatory dynamics -- an inherently nonlinear phenomenon that is abundant in natural and engineering systems such as chemical reactions, electric circuits, structural vibrations, cardiac cells, spiking neurons, and the locomotion of animals and robots~\citep{Winfree1967, Guckenheimer1975, Arnoldbook, Kuramotobook, Ermentroutbook, Strogatz2000, Pikovskybook}. In the context of fluid dynamics, the recent work of Taira and Nakao \cite{Taira2018} was the first to study the synchronization properties of the cylinder flow using phase-reduction analysis, a technique commonly used for biological and chemical systems~\citep{Nakao2016}. 
Periodic forcing continues to present an appealing flow control strategy for a wide range of applications, as it has been shown to effectively reduce bluff body drag~\citep{Pastoor2008}, increase lift of airfoils~\citep{Semaan2016}, and enhance mixing in heat exchangers~\citep{Herrmann2018b}. Understanding synchronization in the context of unsteady aerodynamics is key to leverage periodic flow control and explain the mechanisms that lead to the performance improvements observed in these success stories.

\begin{figure*}[t]
\centering
\includegraphics[width=1\textwidth]{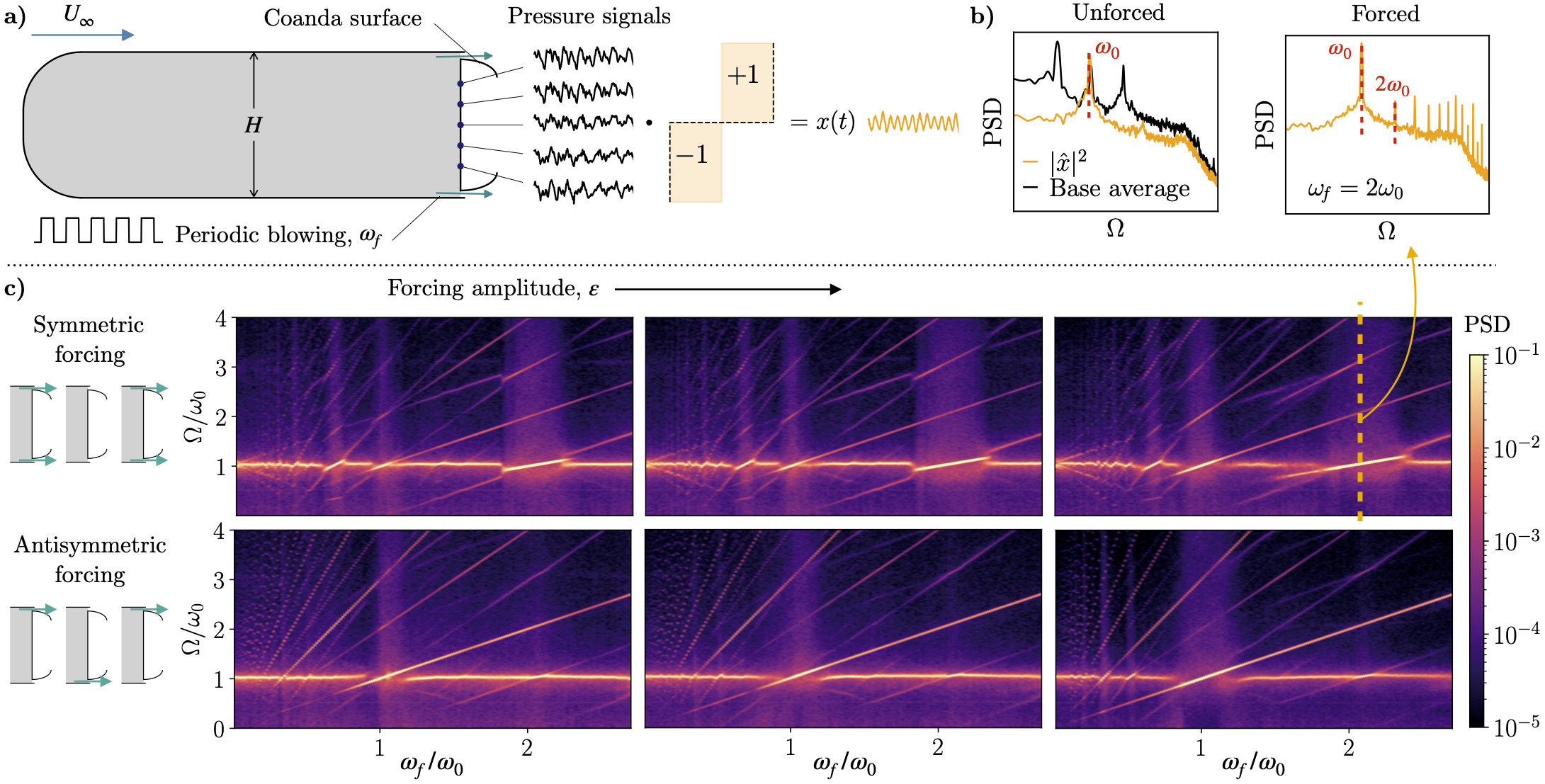}
\vspace{-.3in}
\caption{(\textit{a}) Experimental setup to study the response of the global vortex shedding mode in the wake of a D-shaped bluff body with periodic Coanda blowing at $Re=5.62\times 10^4$ based on the free-stream velocity and body height. Time-resolved pressure measurements are obtained from five sensors located along the mid-span of the rear face of the body. (\textit{b}) Normalized power spectral density (PSD) of the antisymmetric pressure average $x(t)$ that characterizes the global mode amplitude response. (\textit{c}) Panels show the PSD of the time-series of $x(t)$ as a function of excitation frequencies $\omega_f$, for two types of actuation, symmetric and antisymmetric, and three forcing amplitudes $\varepsilon$.}
\label{setup}
\end{figure*}

Recent experimental work by Barros et al. \cite{Barros2016} and Rigas et al. \cite{Rigas2017} reported synchronization in the harmonically forced turbulent wake past an Ahmed body and an axisymmetric blunt body, respectively. Both studies linked the spatial symmetry properties of the forcing mode to the type of response observed. They found the presence of a $1:2$ subharmonic resonance for symmetric disturbances, where the vortex shedding frequency synchronizes to half of the forcing frequency. This lock-on phenomenon was attributed to resonant wave-triads, where the spatial structure of the interacting forcing and response modes are constrained by the triadic consistency condition \citep{Craik1971, Craik1986,Duvvuri2015}. This hypothesis suggests that studying the synchronization properties of forced flows past bluff bodies is a promising way to learn about the dominant nonlinear mode interactions present in the wake.

In previous work, the response of oscillator flows to periodic disturbances has been modeled using the Stuart-Landau equation with the addition of a forcing term \citep{Provansal1987, LeGal2001, Sipp2012, Rigas2017, Boury2018}. In the work of Sipp \cite{Sipp2012}, this model was derived analytically by extending the weakly nonlinear analysis of laminar globally unstable flows to include the effects of external forcing. Rigas et al. \cite{Rigas2017} used an eddy viscosity closure and the phase-averaged Navier-Stokes equations to extend the analysis to the turbulent regime. They validated the resulting model in the proximity of a subharmonic resonance by comparing against experiments of the turbulent wake past an axisymmetric body. Hence, weakly nonlinear analysis \citep{Sipp2012, Rigas2017} provides a theoretically-based structure for a model of the form
\begin{equation}
\td{A}{t} = \sigma A - l|A|^2 A + g(A, F), \ F(t)=F(t+2\pi/ \omega_{\! f}),\label{WNA}
\end{equation}
\noindent
where $g(A,F)$ captures the excitation induced by the coupling between the global mode, with amplitude $A$, and the periodic forcing mode, with amplitude $F$. However, in practice, the computation of $g(A,F)$ requires high-fidelity numerical simulations. Alternatively, recent data-driven techniques are enabling the identification of models directly from data~\citep{Schmid2010,Rowley2009,Mezic2013,Kutz2016book,Brunton2016,Rudy2017,Loiseau2018,Towne2018,Taira2019,Brunton2020}.

Our goal in this work is to obtain new physical insight into the dynamics of forced turbulent oscillator flows by deriving improved mechanistic models from more comprehensive experimental data than has been previously reported. 
As an example of this class of systems, we study the turbulent wake behind a D-shaped bluff body, subject to periodic Coanda blowing, using wind tunnel experimental data, as shown in Fig.~\ref{setup} and described in \S\ref{sec:setup}. In \S\ref{sec:data} we present our experimental dataset characterizing the wake response to periodic forcing for a wide range of excitation parameters. We elucidate the role of resonant triadic interactions for the first time by studying the finely resolved variations of the power spectral density of the global mode response with the excitation frequency. In \S\ref{sec:model} we derive an extended Stuart-Landau model for the evolution of the forced global mode that is amenable to analysis and explain how its coefficients can be identified directly from data. In \S\ref{sec:sync}, we show that our model accurately captures multiple resonances and frequency lock-on regions observed experimentally. Furthermore, model predictions reveal the boundaries of the synchronization regions for the forced D-shaped bluff body wake. Concluding remarks are offered in \S\ref{sec:conclusions}.

\section{Experimental Setup}\label{sec:setup}

A major contribution of this work is the exhaustive experimental investigation of the turbulent wake response to periodic blowing for a broad range of excitation frequencies, excitation amplitudes, and forcing configurations. The experiments are conducted in the ``Leiser Niedriggeschwindigkeitswindkanal Braunschweig" (LNB) wind tunnel at the institute of fluid mechanics of the Technische Universit\"at Braunschweig. The LNB is a continuous atmospheric Eiffel type open return wind tunnel with room recirculation and a closed test section. It has a Burger-type nozzle with a contraction ratio of $16:1$. To reduce turbulence, incoming air is guided through a $30$mm thick fleece mat, a $133$mm thick honeycomb, and a fine woven screen. The resulting turbulence level is below $0.1$\% at $10$m/s. The test section has a width of $400$mm, a height of $600$mm, and a length of $1500$mm. To compensate for boundary layer growth on the walls, the test section has a horizontal opening angle of $1$ degree. The flow is driven by a $9$-blade fan at the end of the diffuser.

The experimental model is a D-shaped bluff body with a blunt trailing edge; it has a height of $53.4$mm, a length of $190.6$mm, and a width of $390$mm. The model is horizontally mounted in the wind tunnel and held by one steel tube on each side. The model nearly spans the entire width of the test section. Zigzag tape is applied to the upper and lower side at about $9$\% body length to trip the boundary layer and to prevent the formation of a laminar separation bubble. A sketch of the model is presented in Fig.~\ref{setup}(\textit{a}). The model is equipped with two Coanda actuators at the trailing corners, each fed by four plenum chambers. The Coanda surfaces have a $9.4$mm radius, which was determined by numerical optimization~\citep{Semaan2018}. The jet slit height is set to $0.2$mm. The uniformity of each jet is verified with a fish mouth probe to be within $10$\% of the mean exit pressure.

Unsteady actuation is enabled through eight Festo MHJ9-QS4-MF monostable 2/2-way valves with an operating pressure range of $0.5$ to $6$bar. The valves can be operated at maximum frequency of $1$kHz. Time-resolved pressure signals are acquired by five Honeywell SLP pressure sensors distributed along the mid-span of the rear face of the model. The sensors have a measurement range of $\pm 1000$Pa differential pressure, a repeatability of $0.5$\% of the full scale, and a response time of $100\mu$s. Plenum pressure is monitored by two Kulite pressure sensors with a range of $\pm 3.5 \cdot 10^4$ Pa differential pressure and an accuracy of $\pm 0.1\%$ of the full scale. All differential pressures are measured relative to the static pressure in the free stream. The instantaneous jet velocity is estimated from the pressure measurements in the plenum chambers. 

\section{Experimental Dataset}\label{sec:data}

The dynamics of turbulent bluff body wakes exhibit self-sustained oscillations. The dominant coherent structure characterizing the oscillatory motion is known as the global vortex shedding mode. We use the antisymmetric average of the five base pressure sensors $x(t)$ to capture the amplitude of the global vortex shedding mode, as shown in Fig.~\ref{setup}(\textit{a}). For the unforced flow, $x(t)$ has a power spectral density with one distinct peak located at the fundamental shedding frequency $\omega_0$, corresponding to a Strouhal number based on the free-stream velocity and body height of $St=\omega_0 H/2\pi U_{\infty}=0.23$, as shown in Fig.~\ref{setup}(\textit{b}). We study the long-term response of the global vortex shedding mode to periodic forcing, for a broad range of parameters, including the forcing configuration, and the excitation amplitude and frequency. Two forcing configurations are considered: in-phase blowing through the top and bottom slits, leading to the excitation of a spatially symmetric flow structure, and $180^\circ$ out-of-phase blowing, exciting a spatially antisymmetric structure. The effect of forcing amplitude is investigated for three blowing intensities quantified non-dimensionally by the  momentum coefficient as
\begin{equation}
c_{\mu}=2\frac{h}{H}\frac{U^2_{\text{jet}}}{U_{\infty}^2},
\end{equation}
\noindent
where $U_{\text{jet}}$ is the root mean square value of the jet velocity calculated from the plenum pressure measurements, $h=0.2$mm is the slot height, $H=53.4$mm is the body height, and the factor $2$ accounts for the number of actuators. We define the excitation amplitude $\varepsilon$ as the average of $c_{\mu}$ over \emph{all} excitation frequencies at a given tank pressure.

For each forcing configuration and amplitude, the base pressure is recorded for $8$s at a sampling rate of $5$kHz for blowing frequencies between $1$Hz and $210$Hz with $1$Hz intervals. For all cases, the conditions are maintained for $5$s to ensure steady state behavior before the measurements are taken. The complete dataset is conformed by a total of $1260$ time-series of $x(t)$, each consisting of $40000$ samples. These time-series are used to characterize the mean frequency and amplitude of the wake response. Each time-series $x(t)$ is low-pass filtered using a $5^{\text{th}}$-order Butterworth filter with a cutoff frequency of $1.3\omega_0$. Its Hilbert transform $x_H(t)$ is then computed to build an analytic signal for the complex global mode amplitude $A(t)=(x+\ii x_H)(t)$, which is a common practice when studying oscillatory dynamics from data \citep{Pikovskybook}. 
Once we have the complex time-series $A(t)$, we compute its mean amplitude $r$, and its mean frequency $\omega$ from the mean of the time derivative of its instantaneous phase.

\begin{figure*}[tb]
\centering
\vspace{-.15in}
\includegraphics[width=1\textwidth]{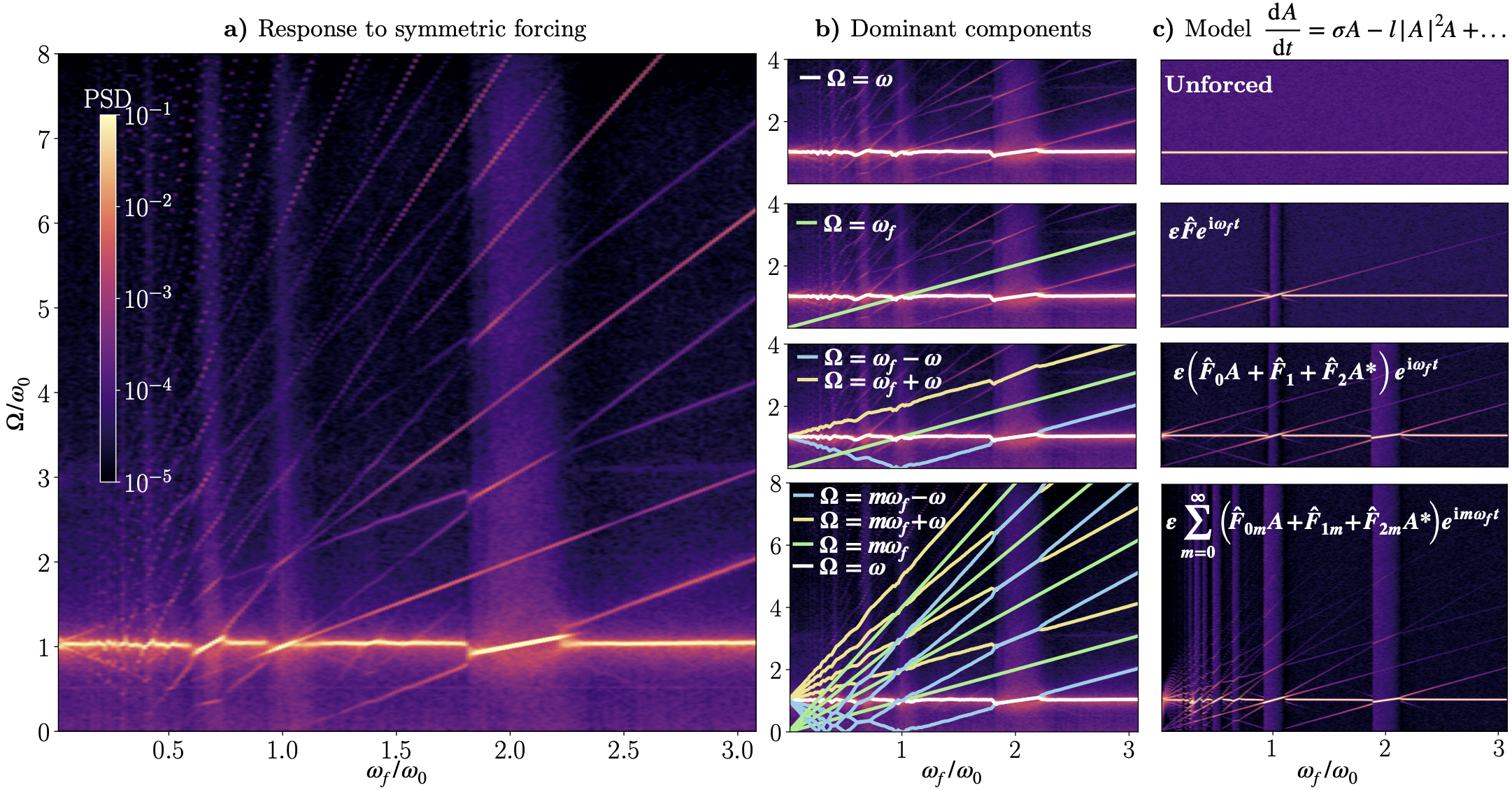}
\vspace{-.3in}
\caption{Experimental response of the global vortex shedding mode amplitude in the turbulent wake of a D-shaped bluff body under periodic symmetric forcing characterized via its PSD as function of the excitation frequency -- \emph{the power spectrum response} (a).  The interpretation of the dominant components is shown in (\textit{b}), and (\textit{c}) shows the power spectrum response computed from simulations of a hierarchy of modified Stuart-Landau models. White measurement noise is added to all models.}
\label{spectrum_responseS}
\vspace{-.1in}
\end{figure*}

The mean wake response provides information about resonances and lock-on regions that has already been discussed is in previous studies \citep{Barros2016, Rigas2017}. A deeper insight into the underlying nonlinear interactions is obtained by examining the power spectral density (PSD) of the global mode response at different forcing frequencies. The normalized PSD of $x(t)$ is computed using Welch's method \citep{Welch1967}, splitting the time-series into $10$ segments with $50\%$ overlap and tapered by a Hanning window. The resulting PSD obtained for each excitation frequency are stacked as columns in a heat map, shown in Fig.~\ref{setup}(\textit{c}). The large number of test cases yields a fine resolution of the effect of  $\omega_{\! f}$ on the PSD of the global vortex shedding mode, resulting in a highly interpretable visualization that we refer to as the power spectrum response.

An enlarged version of the power spectrum response for the case of symmetric forcing is shown in Fig.~\ref{spectrum_responseS}(\textit{a}). Identification of the dominant frequency components of the response, $\Omega$, provides insight into the nature of the interactions between forcing and response modes. The highest energy component coincides with the observed mean vortex shedding rate, $\Omega=\omega$. In addition, Fig.~\ref{spectrum_responseS}(\textit{a}) displays components following sharp straight lines, each with a physical interpretation, as shown in the panels of \ref{spectrum_responseS}(\textit{b}). As expected, the response exhibits a strong component at the excitation frequency $\Omega=\omega_{\! f}$. Moreover, triadic interactions are readily seen in this visualization as side-bands to the forcing frequency component, i.e., $\Omega=\omega_{\! f} \pm \omega$. All other relevant components of the power spectrum response for this case are explained as higher harmonics of the forcing and the triadic interactions with these higher harmonics. Fig.~\ref{spectrum_responseS}(\textit{c}) shows the result of our low order model, which we develop in the next section.

\begin{figure*}[tb]
\centering
\vspace{-.15in}
\includegraphics[width=1\textwidth]{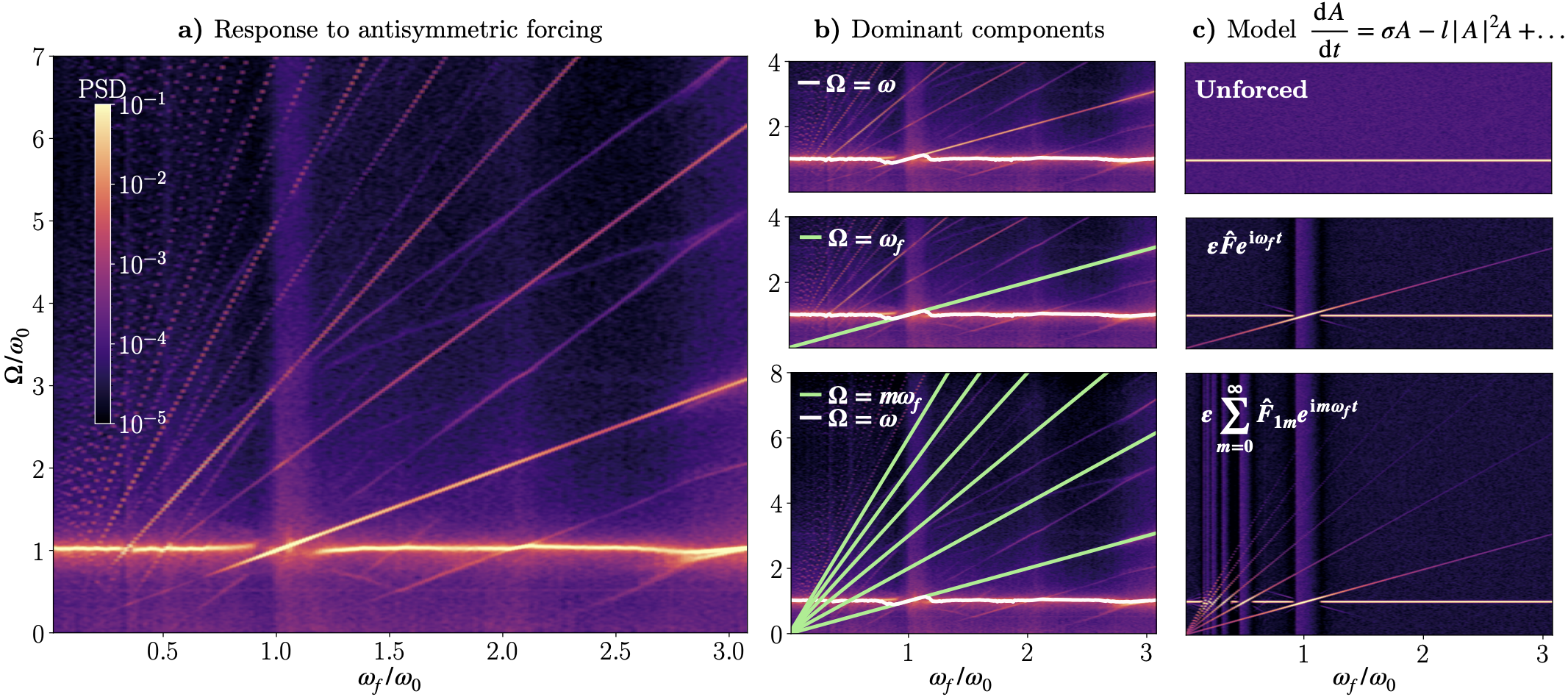}
\vspace{-.3in}
\caption{Experimental response of the global vortex shedding mode amplitude in the turbulent wake of a D-shaped bluff body under periodic antisymmetric forcing characterized via its PSD as function of the excitation frequency -- \emph{the power spectrum response} (a). The interpretation of the dominant components is shown in (\textit{b}), and (\textit{c}) shows the power spectrum response computed from simulations of a hierarchy of modified Stuart-Landau models. White measurement noise is added to all models.}
\label{spectrum_responseA}
\vspace{-.1in}
\end{figure*}

For antisymmetric forcing, the power spectrum response is quite different, as shown in Fig.~\ref{spectrum_responseA}(\textit{a}). In this case, the dominant frequency components in the power spectrum response correspond to the vortex shedding rate $\Omega=\omega$ and to harmonics of the forcing $\Omega=m\omega_{\! f}$, with $m$ being an integer, as shown in Fig.~\ref{spectrum_responseA}(\textit{b}). The absence of components $m\omega_{\! f} \pm \omega$ induced by nonlinear interactions is expected because the forcing mode and the pair of conjugate response modes have spatial symmetries that are incompatible with the triadic consistency condition, i.e. their wavenumbers do not sum to zero. The other present frequency components, correspond to interactions between the forcing and higher harmonics of the response, which are orders of magnitude weaker in power and do not play a significant role in the wake dynamics.

\section{Modified Stuart-Landau Model}\label{sec:model}

We now summarize the derivation of our mechanistic model, its use to analyze the dynamics of the system, and how to obtain its coefficients directly from data.

\subsection{Derivation of the Model}

We derive a parsimonious model for the evolution of the amplitude of the global vortex shedding mode in the turbulent wake behind a bluff body. We begin with the forced Stuart-Landau model in \eqref{WNA} obtained from weakly nonlinear theory~\citep{Sipp2007}. Our goal is to obtain an interpretable model for the excitation function $g(A, F)$ that explains the interactions with the periodic forcing $F(t)=F(t+2\pi/ \omega_{\! f})$. We posit a simple structure for $g$ and proceed by systematically increasing its complexity until the resulting power spectrum response exhibits the same features observed in Fig.~\ref{spectrum_responseS}(\textit{b}). For easier comparison we evaluate each model in this hierarchy at the same actuation frequencies as our experimental data using the same number of sample records with the same length and sampling rate. Measurement noise is added to the numerical solutions before computing the power spectrum response to represent background turbulence \citep{Semaan2016}. Zero-mean white Gaussian noise is used with a standard deviation proportional to the corresponding response amplitude for each $\omega_{\! f}$. This enables unbiased comparison between the model response and experimental data, as in Fig.~\ref{spectrum_responseS}(\textit{c}) and Fig.~\ref{spectrum_responseS}(\textit{a}), respectively.

We first show the power spectrum response in the absence of forcing $(g=0)$. As expected, the spectral power is localized at the natural shedding frequency, as shown in the first panel of Fig.~\ref{spectrum_responseS}(\textit{c}). Next, we consider the case of harmonic forcing at the known actuation frequency, $F=\varepsilon\hat{F} e^{\ii \omega_{\! f} t}$, where $\varepsilon$ is the forcing amplitude and $\hat{F}$ is a constant that depends on the spatial structure being excited in the flow. This harmonic forcing acts linearly on the dynamics, i.e. $g=F$, as shown in the second panel in Fig.~\ref{spectrum_responseS}(\textit{c}). As a third model in this hierarchy, we add terms involving quadratic interactions of the forcing mode with the global mode and with its complex conjugate, $g=AF_o + F_1 + A^* F_2$, as shown in the third panel in Fig.~\ref{spectrum_responseS}(\textit{c}), where we allow for the possibility that different flow structures $\hat{F}_i$ are excited by each term. Finally, we allow for non-harmonic external forcing, which can be expanded as a Fourier series
\begin{equation}
F_i(t)=\varepsilon\hspace{-0.25cm}\sum_{m=-\infty}^{\infty} \hspace{-0.25cm}\hat{F}_{im} e^{\ii m\omega_{\! f} t},
\end{equation}
where $\hat{F}_{im}$ depends on the spatial structure being excited in the flow by the $m^{\text{th}}$ harmonic of the forcing. The resulting model for the forced complex amplitude of the global mode is a modified Stuart-Landau equation:
\begin{equation}
\td{A}{t} \!=\! \sigma \! A \! -\!  l|\! A|^2 \! A \! +\! \varepsilon\hspace{-0.28cm}\sum_{m=-\infty}^{\infty} \hspace{-0.25cm}\left(\! \hat{F}_{0m} A \! + \! \hat{F}_{1m} \! +\! \hat{F}_{2m} A^*\! \right)\!  e^{\ii m\omega_{\! f} t}. \label{A}
\end{equation}

The coefficients $\hat{F}_{0m}, \ \hat{F}_{1m}$ and $\hat{F}_{2m}$, for integer $m$, parametrize the direct and parametric excitation terms induced by the periodic forcing. These differ depending on the forcing configuration, symmetric or antisymmetric, and are identified from data based on the analysis presented in the following section. The model power spectral response remarkably resembles the results obtained from the turbulent flow experiments, as shown in the last panel in Fig.~\ref{spectrum_responseS}(\textit{c}). When carrying out the analogous procedure for antisymmetric forcing, we find that the same model generalizes to both forcing modes. Nevertheless, in the latter case the terms $\hat{F}_{0m}$ and $\hat{F}_{2m}$ are negligible due to the absence of triadic interactions, as shown in Fig.~\ref{spectrum_responseA}(\textit{c}). 
In the next section, we will show that under nearly resonant forcing, our extended Stuart-Landau model simplifies to the  model of Rigas et al. \cite{Rigas2017}, generalized to non-harmonic excitation. In addition, our model also explains the response observed away from resonances in our extensive experimental data set.

\subsection{Model Based Analysis}

We now demonstrate the value of our modified Stuart-Landau model. We exploit the fact that the dynamical system given by \eqref{A} is amenable to classical analysis of nonlinear oscillators. When there is no actuation, the solution to \eqref{A} for $A=r e^{\ii\theta}$ exhibits an unstable fixed point at $r=0$ and a stable limit cycle of radius $r_0=\sqrt{\sigma_r/ l_r}$ with frequency $\omega_0=\sigma_i - l_i r_0^2$; the subscripts $r$ and $i$ denote the real and imaginary parts of the complex constants. In the presence of periodic excitation, we are interested in finding frequency locked solutions, where the phase of the complex amplitude rotates with a mean angular frequency $\langle\dot{\theta}\rangle=\omega$, which is a rational multiple of the forcing frequency $\omega_{\! f}$. For this purpose, we introduce the change of variables $A=r e^{\ii \theta}=r e^{\ii (\phi + \omega t)} = \tilde{A}e^{\ii \omega t}$, where $\tilde{A}=r e^{\ii \phi}$ has a slowly varying amplitude $r$ and phase $\phi$. Substituting into \eqref{A} and rearranging yields
\begin{multline}
\left[\td{\tilde{A}}{t} -  (\sigma-\ii \omega) \tilde{A} + l|\tilde{A}|^2\tilde{A}\right]e^{\ii \omega t} =
\\
\varepsilon \hspace{-0.25cm} \sum_{m=-\infty}^{\infty}\hspace{-0.25cm}  \left(\hat{F}_{0m} \tilde{A} e^{\ii \omega t}+ \hat{F}_{1m}+\hat{F}_{2m} \tilde{A}^*e^{-\ii\omega t}\right)e^{\ii m\omega_{\! f}t},\label{A_tilde}
\end{multline}
\noindent
which describes the evolution of the slow complex amplitude $\tilde{A}$. Up to this point, no approximations have been made. We have managed to turn the search for frequency locked solutions of $A$ into the search for fixed points of $\tilde{A}$. Nevertheless, \eqref{A_tilde} is non-autonomous, meaning that the dynamics depend explicitly on $t$. By applying the method of averaging \citep{GyHbook}, also known as the Krylov-Bogoliubov method, we can further simplify this equation to an autonomous dynamical system. In practice this is achieved by integrating over one period of the complex amplitude $T=2\pi/\omega$ and neglecting the changes of the slow variables over that time horizon. The terms on the right hand side of \eqref{A_tilde} that are kept after the averaging procedure are those that cancel out the fast oscillations $\sim e^{\ii \omega t}$ on the left. Therefore, the resulting expression for the slow dynamics depends on the actuation frequency. The three possible cases are analyzed below: non resonant forcing, harmonic resonance $n \omega_{\! f} \approx \omega$, and subharmonic resonance $n \omega_{\! f} \approx 2 \omega$, for integer $n$.


Away from any resonance of the system, $n \omega_{\! f}  \ne \omega$ and $n \omega_{\! f}  \ne 2\omega$, the slow dynamics are governed by the balance between the left hand side of \eqref{A_tilde} and the zeroth harmonic of $F_0$, as follows
\begin{equation}
\td{\tilde{A}}{t} = (\sigma-\ii \omega) \tilde{A} - l|\tilde{A}|^2 \tilde{A} + \varepsilon\hat{F}_{00} \tilde{A}.
\end{equation}
\noindent
Notice that $\hat{F}_{00}$ changes the eigenvalue of the linear dynamics of the global mode at the origin to $\sigma'=\sigma + \varepsilon\hat{F}_{00}$. As a consequence, the radius and frequency of the stable limit cycle are modified according to
\begin{equation}
r_0'^2=r_0^2+\varepsilon\hat{F}_{00}/ l_r \quad \text{and} \quad  \omega'_0=\omega_0 - \varepsilon\hat{F}_{00} l_i/ l_r.\label{rp}
\end{equation}
\noindent
After identifying $\sigma$ and $l$ from the unforced dynamics, we can solve for $\hat{F}_{00}$ from \eqref{rp} using measurements of the long-term unforced response, $r_0$ and $\omega_0$, along with measurements of the modified response away from resonances, $r_0'$ and $\omega_0'$. The values identified for $\hat{F}_{00}$ in the cases of symmetric and antisymmetric forcing of the turbulent D-shaped bluff body wake are shown in Tab.~\ref{parameters}.


In the proximity of an order $n$ harmonic resonance, i.e., $n \omega_{\! f} = \omega$, the slow dynamics are governed by the balance between the left hand side of \eqref{A_tilde} and the terms on the right hand side that include $\hat{F}_{00}$,  $\hat{F}_{1n}$, and $\hat{F}_{22n}$. Furthermore, if the linear interaction of the global mode with the $n^{\text{th}}$ harmonic of the forcing dominates over the nonlinear interaction with its $2n^{\text{th}}$ harmonic, then $\hat{F}_{1n} \gg \hat{F}_{22n}\tilde{A}^*$, resulting in
\begin{equation}
\td{\tilde{A}}{t} = (\sigma'-\ii \omega) \tilde{A} - l|\tilde{A}|^2 \tilde{A} + \varepsilon\hat{F}_{1n},
\label{F1}
\end{equation}
where $\sigma'$ includes the contribution of the interaction with $F_0$. Recasting the system in polar form we obtain
\begin{subequations}
\begin{equation}
\td{r}{t} =  -l_r r \left(r^2 - r_0'^2  \right) + \varepsilon\hat{F}_{1n} \cos(\phi),
\end{equation}
\begin{equation}
r\td{\phi}{t} = r\left(\omega_0' - n\omega_{\! f}\right) - l_i r \left(r^2 - r_0'^2  \right) - \varepsilon\hat{F}_{1n} \sin(\phi).
\end{equation}
\label{F1_polar}
\end{subequations}
This system of equations has no explicit time dependence, hence we can search for fixed points by setting $\td{r}{t}=\td{\phi}{t}=0$, which represent frequency locked solutions of the form $A=re^{\ii n \omega_{\! f} t}$. Furthermore, equating the $\left(r^2 - r_0'^2  \right) $ terms and rearranging yields
\begin{equation}
\omega_0' - n\omega_{\! f} = \frac{\varepsilon\hat{F}_{1n}}{r}\left(\sin(\phi)+\cos(\phi)\frac{l_i}{l_r} \right).
\end{equation}
\noindent
Thus, a frequency locked solution exists only if the frequency detuning is in the range
\begin{equation}
|\omega_0' - n\omega_{\! f} | < \frac{\varepsilon\hat{F}_{1n}}{r_0}\sqrt{1+\left(\frac{l_i}{l_r}\right)^2} ,
\label{sync1}
\end{equation}
where $r_0$ approximates $r$ evaluated at the synchronization boundary. Therefore, \eqref{sync1} represents the bounds for the $n:1$ synchronization region. Furthermore, having already identified $l$ from the unforced dynamics, this expression can be used to identify the coefficients $\hat{F}_{1n}$ from data. This is achieved using experimental measurements of the mean global mode frequency $\omega$, and finding the values of $\omega_{\! f}$ that delimit the corresponding $n:1$ frequency lock-on regions, i.e., where $|\omega - n\omega_{\! f}|<\delta$ is satisfied, with $\delta$ being a small threshold. Subsequently, we subtract the upper and lower bounds to compute the width of the frequency lock-on regions and substitute them into \eqref{sync1}, allowing the computation of one of these model parameters for every harmonic resonance considered. In the present experiment we observe three harmonic resonances for both symmetric and antisymmetric forcing. The identified coefficients are presented in Tab.~\ref{parameters}.


We now investigate the cases where a subharmonic resonance of the type $n \omega_{\! f} =2 \omega$ takes place, excluding even values of $n$ as those were accounted for in the previous scenario.
In the proximity of a subharmonic resonance, the slow dynamics are governed by the balance between the left hand side of \eqref{A_tilde} and the terms on the right hand side that include $\hat{F}_{00}$,  and  $\hat{F}_{2n}$, as follows
\begin{equation}
\td{\tilde{A}}{t} = (\sigma'-\ii \omega) \tilde{A} - l|\tilde{A}|^2 \tilde{A} + \varepsilon\hat{F}_{2n}\tilde{A}^*,
\label{F2}
\end{equation}
\noindent
where $\sigma'$ includes the contribution of the interaction with $F_0$. Recasting \eqref{F2} in polar form we obtain
\begin{align*}
\td{r}{t} &=  -l_r r \left(r^2 - r_0'^2  \right) + \varepsilon\hat{F}_{2n} r\cos(2\phi),\\
r\td{\phi}{t} &= r\left(\omega_0' - n\omega_{\! f}/2 \right) - l_i r \left(r^2 - r_0'^2  \right) - \varepsilon\hat{F}_{2n} r\sin(2\phi).
\end{align*}
\label{F2_polar}

As in the previous case, these equations have no explicit time dependence, hence we search for fixed points by setting $\td{r}{t}=\td{\phi}{t}=0$, which represent frequency locked solutions that are now of the form $A=re^{\ii n \omega_{\! f}/2 t}$. Again, equating the $\left(r^2 - r_0'^2  \right) $ terms and rearranging yields
\begin{equation}
\omega_0' - n\omega_{\! f}/2 = \varepsilon\hat{F}_{2n}\left(\sin(2\phi)+\cos(2\phi)\frac{l_i}{l_r} \right),
\end{equation}
\noindent
for which a solution exists only for the range of the frequency detuning given by
\begin{equation}
|\omega_0' - n\omega_{\! f}/2| < \varepsilon\hat{F}_{2n}\sqrt{1+\left(\frac{l_i}{l_r}\right)^2}.
\label{sync2}
\end{equation}
This inequality determines the bounds for the $n:2$ synchronization region. Moreover, having already identified $l$ from the unforced dynamics, this expression can be used to identify the coefficients $\hat{F}_{2n}$ from data. Similar to the case of harmonic resonance, this is achieved using experimental measurements of the mean global mode frequency $\omega$, and finding the values of $\omega_{\! f}$ that delimit the corresponding $n:2$ frequency lock-on regions, i.e., where $|\omega - n\omega_{\! f}/2|<\delta$ is satisfied, with $\delta$ being a small threshold. Subsequently, we calculate the width of the frequency lock-on regions and substitute them into \eqref{sync2}, allowing the computation of one of these model parameters for every subharmonic resonance observed. For the present flow, we observe four subharmonic resonances for symmetric forcing, and one for antisymmetric. The identified coefficients are shown in Tab.~\ref{parameters}.

\subsection{Identifying the Stuart-Landau Coefficients}

\begin{figure}[t]
\centering
\includegraphics[width=0.8\columnwidth]{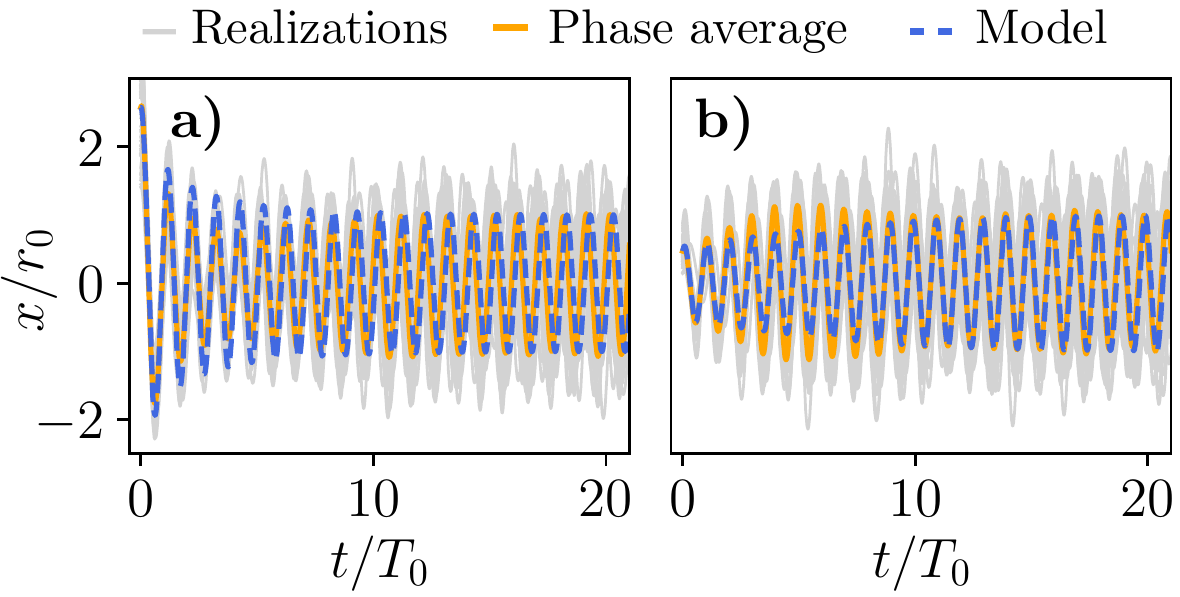}
\vspace{-.15in}
\caption{Transient evolution of the global vortex shedding mode amplitude in the unforced flow past a D-shaped bluff body starting from: (\textit{a}) the subharmonic resonant response to periodic forcing, or (\textit{b}) the long-term response to steady blowing. Gray curves show the superposition of all experimental realizations, orange curves show their phase-average, and the dashed blue curves correspond to the identified model.}
\label{transients}
\end{figure}

In the absence of forcing, the proposed model, given by \eqref{A}, reduces to the classic Stuart-Landau equation that is parametrized by the complex coefficients $\sigma,$ and $l$. Therefore, these unknown parameters can be identified using measurements of transients of the system when the forcing is not active. To characterize the unforced dynamics, we drive the system away from its long-term behavior using steady blowing and using resonant periodic forcing. Starting from each of these conditions, the forcing is turned off and we record the evolution of the global mode. These transients are low-pass filtered using a $5^{\text{th}}$-order Butterworth filter with a cutoff frequency of $1.3\omega_0$ and then phase-averaged over an ensemble of $59$ realizations. Figures \ref{transients}(\textit{a}) and \ref{transients}(\textit{b}) show the evolution of the antisymmetric pressure average starting from larger and smaller amplitude oscillations, respectively. In both cases it takes on the order of $~10$ shedding cycles for the wake to relax onto the limit cycle describing its long-term behavior.

Based on the data for the unforced transients, we identify $\sigma$ and $l$ through a constrained least squares regression. The phase-averaged and filtered time-series $x(t)$ and its Hilbert transform $x_H(t)$ are used to build an analytic signal for the complex global mode amplitude ${A(t)=(x+\ii x_H)(t)}$. Once we have the instantaneous complex amplitude, we can compute its instantaneous amplitude $r(t)$ and phase $\theta(t)$, as well as the respective time derivatives using $2^\text{nd}$-order central finite differences. In addition to the transients, we measure the mean amplitude and frequency, $r_0$ $\omega_0$, from the magnitude and position of the peak of the PSD of the long-term unforced time-series. In accordance with the Stuart-Landau equation, we know that these measurements must satisfy the solutions for the radius $r_0=\sqrt{\sigma_r/ l_r}$ and frequency $\omega_0=\sigma_i - l_i r_0^2$ of the stable limit cycle, where the subscripts $r$ and $i$ denote the real and imaginary parts of the complex constants. Therefore, the regression problem is formulated by recasting the Stuart-Landau equation in polar form and deriving linear equality constraints from the known limit cycle solution, as follows
\begin{multline*}
\underbrace{\left[\begin{array}{c}
\dot{r}\\
r\dot{\theta}
\end{array}\right]}_{\b{b}}
=
\underbrace{\left[\begin{array}{cccc}
r & -r^3 & 0 & 0\\
0 & 0 & r & -r^3
\end{array}\right]}_{\b{\Theta}}
\underbrace{\left[\begin{array}{c}
\sigma_r\\
l_r\\
\sigma_i\\
l_i
\end{array}\right]}_{\b{\xi}},
\\
\text{subject to} \quad
\underbrace{\left[\begin{array}{cccc}
1 & -r_0^2 & 0 & 0\\
0 & 0 & 1 & -r_0^2
\end{array}\right]}_{\b{C}}
\underbrace{\left[\begin{array}{c}
\sigma_r\\
l_r\\
\sigma_i\\
l_i
\end{array}\right]}_{\b{\xi}}
=
\underbrace{\left[\begin{array}{c}
0\\
\omega_0
\end{array}\right]}_{\b{d}}.\label{regression}
\end{multline*}
Here, the vector $\b{b}$ and the matrix $\b{\Theta}$ are constructed by vertically stacking the time-series data for $r, \dot{r},$ and $\dot{\theta}$ and using null vectors of the same length. The matrix $\b{C}$ and the vector $\b{d}$ are also computed from experimental data, since we have measured $r_0$ and $\omega_0$. Therefore, the coefficients $\b{\xi}$ can be identified from
\begin{equation}
\min_{\b{\xi}} ||\b{\Theta}\b{\xi}-\b{b}||_2^2, \quad \text{subject to} \quad \b{C}\b{\xi}=\b{d},
\end{equation}
\noindent
which is a convex optimization problem that can be solved using readily available software; we use CVXPY developed by Diamond and Boyd \cite{Diamond2016}. The identified model is simulated and compared against the experimental data in Fig.~\ref{transients}. The identified values for $\sigma$ and $l$ are shown in Tab.~\ref{parameters}.

\begin{table}[t]
\centering
\caption{Identified model coefficients.}
\begin{ruledtabular} 
\begin{tabular}{c d c d c d}
\multicolumn{2}{c}{Unforced} & \multicolumn{2}{c}{Antisymmetric} & \multicolumn{2}{c}{Symmetric}\\
\multicolumn{2}{c}{dynamics} & \multicolumn{2}{c}{forcing} & \multicolumn{2}{c}{forcing}\\
\hline
$\sigma_r$ & $0.028$ & $\hat{F}_{00}$ & $-1.507$ & $\hat{F}_{00}$ & $-1.709$\\
$\sigma_i$ & $1.483$ & $\hat{F}_{11}$ &  $0.383$ & $\hat{F}_{11}$ &  $0.178$\\
$l_r$ & $19.281$ & $\hat{F}_{12}$ & $0.091$ & $\hat{F}_{12}$ & $0.038$\\
$l_i$ & $37.068$ & $\hat{F}_{13}$ & $0.059$ & $\hat{F}_{13}$ & $0.052$\\
 &  & $\hat{F}_{21}$ & $1.057$ & $\hat{F}_{21}$ & $9.099$\\
 &  &  &  & $\hat{F}_{23}$& $5.240$\\
 &  &  &  & $\hat{F}_{25}$ & $2.252$\\
 &  &  &  & $\hat{F}_{27}$ & $0.627$\\
\end{tabular}\label{parameters}
\end{ruledtabular}
\end{table}

\section{Results and Discussion}\label{sec:sync}

\begin{figure}[t]
\centering
\includegraphics[width=1\columnwidth]{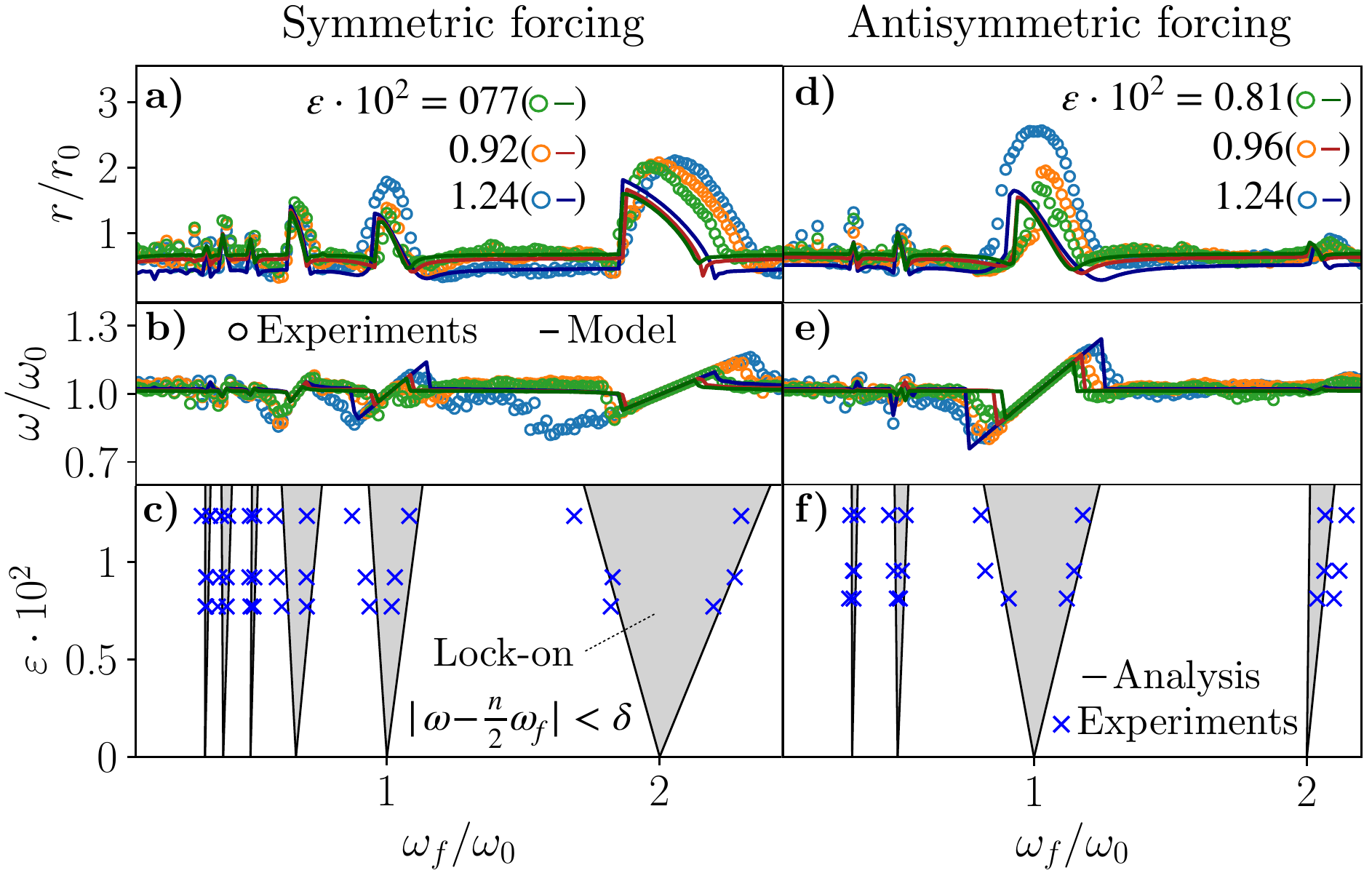}
\vspace{-.25in}
\caption{Response of the global vortex shedding mode amplitude in the turbulent and periodically forced wake of a D-shaped bluff body characterized via its mean amplitude $r$ in (\textit{a}) and (\textit{d}) and mean frequency $\omega$ in (\textit{b}) and (\textit{e}) as a function of the excitation frequency $\omega_f$ for three forcing amplitudes $\varepsilon$. Markers denote experimental results and solid lines correspond to the proposed model. The Arnold tongues, i.e., regions in $\omega_{\! f}$--$\varepsilon$ space where the wake response synchronizes to a rational multiple of the excitation frequency, are shown in (\textit{c}) and (\textit{f}). Experimentally observed boundaries of frequency locked regions are shown by the blue markers. Shaded regions correspond to the synchronization criteria according to \eqref{sync1} and \eqref{sync2}. Panels (\textit{a})-(\textit{c}) and (\textit{d})-(\textit{f}) correspond to symmetric and antisymmetric forcing, respectively.}
\label{Tongues}
\end{figure}

In this section we present a comparison between predictions from our proposed model and experimental data, and discuss the synchronization properties of the forced turbulent wake past a D-shaped body. Using our model with the coefficients shown in Tab.~\ref{parameters}, we compute the long-term amplitude and frequency response of the global vortex shedding mode under periodic forcing as a function of the excitation frequency. A comparison against experimental data for symmetric and antisymmetric forcing at three excitation amplitudes is shown in Figs.~\ref{Tongues}(\textit{a}), (\textit{b}), (\textit{d}),  and (\textit{e}). The modified Stuart-Landau model captures the experimental behavior very well, displaying multiple resonances and frequency lock-on regions. Moreover, the same model structure allows for the description of the long-term response with various forcing amplitudes and different forcing configuration, which translates into different dominating resonances; harmonic or subharmonic.

In accordance with the work of Barros et al. \cite{Barros2016}, when the forcing is symmetric, the largest amplification of the response occurs when the system is excited near twice the fundamental frequency $\omega_{\! f}/\omega_0 \approx 2$, as shown in Fig.~\ref{Tongues}(\textit{a}). As pointed out by Rigas et al. \cite{Rigas2017}, this subharmonic resonance arises due to triadic interactions between the global mode, its complex conjugate and the forcing mode. This can only occur when the spatial characteristics of these modes satisfy the triadic consistency conditions, meaning that, if expressed using a travelling wave ansatz, their wavenumbers and frequencies must add up to zero \citep{Craik1971, Craik1986}. This requirement is clearly not satisfied for the case of antisymmetric forcing, where the largest amplification of the response occurs at the harmonic resonance $\omega_{\! f}/\omega_0 \approx 1$, as shown in Fig.~\ref{Tongues}(\textit{d}).

We now investigate which combinations of the actuation frequency and amplitude cause the turbulent wake to synchronize with the external excitation. These regions in $\omega_{\! f}$--$\varepsilon$ space are known as Arnold tongues \citep{Arnoldbook}. From experimental data, the mean frequency of the response $\omega$ is used to find the excitation frequencies that delimit the $n:1$ harmonic and $n:2$ subharmonic lock-on regions. These are marked with blue crosses in Figs.~\ref{Tongues}(\textit{c}) and (\textit{f}) for both symmetric and antisymmetric forcing, respectively. The model based Arnold tongues which are shaded in Figs.~\ref{Tongues}(\textit{c}) and (\textit{f}) correspond to the regions bounded by ~\eqref{sync1} and \eqref{sync2}. As the figure shows, our modified Stuart-Landau model captures the observed tongues and predicts the position of the transition boundaries for the range of excitation amplitudes studied. In this particular example, significant drag reduction is observed outside the Arnold tongues, which highlights the relevance of modeling synchronization as a key enabler of effective periodic flow control.

\section{Conclusions}\label{sec:conclusions}

In this work, we have leveraged a uniquely comprehensive experimental dataset of a periodically forced turbulent wake to develop a modified Stuart-Landau model for the response of the global vortex shedding mode. The breadth and quality of our dataset reveals previously unobserved resonances and frequency lock-on regions. Moreover, it enables the construction of heat maps showing the power spectral density of the wake response as a function of the excitation frequency. This novel visualization exposes triadic interactions as dominant frequency components, providing the first confirmation of their previously conjectured role in the forced wake response \cite{Barros2016, Rigas2017}.

We derive a parsimonious model for the evolution of the forced global mode that generalizes previous models to non-harmonic excitation and extends their applicability to a broader range of parameters. From this model we gain new insights into the nonlinear mechanisms underlying synchronization of the flow response to external excitation. We find excellent agreement between the model and experiments, building a foundation for future investigations into effective periodic flow control. Low-order models that capture synchronization mechanisms are essential for the design of feedback controllers to manipulate periodic coherent structures that govern lift, drag and mixing in oscillator flows. Another key advantage of the model is that its unknown coefficients can be identified directly from data. Our approach to expose the synchronization properties of the system in an extensive parameter space using only a few measurements is promising, as it can be generalized to a large class of forced oscillator flows.

\section*{Acknowledgments}
\vspace{-.1in}
This work has been supported by the PRIME programme of the German Academic Exchange Service (DAAD) with funds from the German Federal Ministry of Education and Research (BMBF) and by the Deutsche Forschungsgemeinschaft (DFG) project number SE 2504/3-1. 
SLB acknowledges funding support from the Air Force Office of Scientific Research (AFOSR FA9550-18-1-0200) and the Army Research Office (ARO W911NF-19-1-0045).

\bibliography{library}

\end{document}